**Discrimination of virgin olive oils with different geographical origin:**

**a rapid untargeted chromatographic approach based on volatile compounds**

Rosa Palagano[a], Enrico Valli[a], Chiara Cevoli[a], Alessandra Bendini[a,*], Tullia Gallina Toschi[a]

[a] Department of Agricultural and Food Sciences (DISTAL), University of Bologna, Cesena, 47521, Italy

* Corresponding author: alessandra.bendini@unibo.it (Alessandra Bendini), Department of Agricultural and Food Sciences (DISTAL) - University of Bologna, Piazza Goidanich 60, Cesena (FC), 47521, Italy, Tel +39 0547 338121.

**Abstract**

Many studies have shown that the geographic origin is one of the most influencing factors in consumers' choice of olive oil. To avoid misleading, European regulation has established specific rules to report the geographical origin of extra virgin (EVOOs) and virgin olive oils (VOOs) on the product label, but an official analytical procedure to verify this information has not been yet defined. In this work, a flash gas chromatography untargeted approach for determination of volatile compounds, followed by a chemometric data elaboration, is proposed for discrimination of EVOOs and VOOs according to their geographical origin (EU and Extra-EU). A set of 210 samples was analyzed and two different classification techniques were used, one linear (Partial Least Square-Discriminant Analysis, PLS-DA) and one non-linear (Artificial Neural Network, ANN). The two models were also validated using an external data set. Satisfactory results were obtained for both chemometric approaches: considering the PLS-DA, 89% and 81% of EU and Extra-EU samples, respectively, were correctly classified; for ANN the percentages were 93% and 89%, respectively. These results confirm the reliability of the method as a rapid approach to discriminate EVOOs and VOOs according to their geographical provenance.

**Key words**

Virgin olive oil; Geographical origin; PLS-DA; Untargeted approach; Volatile compounds; ANN.

**Abbreviations**

ANN: Artificial Neural Network; EVOO: Extra Virgin Olive Oil; FGC: Flash Gas Chromatography; PLS-DA: Partial Least Square - Discriminant Analysis; VOO: Virgin Olive Oil.

# 1. Introduction

Over the last 40 years many investigations have been focused on understanding what attributes are important determinants in consumer choice, which have highlighted that the geographic origin is one of the most influencing factors for olive oil (Dekhili, Sirieix, & Cohen, 2011; Del Giudice, Cavallo, Caracciolo, & Cicia, 2015).

In order to ensure that consumers are not misled, the fourth article of the EU Reg. 29/2012 establishes that *"Extra virgin and virgin olive oil shall bear a designation of origin on the labelling"*. This means that for extra virgin (EVOOs) and virgin olive oils (VOOs) commercialized within the EU, it is mandatory to specify the geographical provenance on the label of the product following specific rules. If an oil comes from an EU Member State or third country, a reference to the EU Member State, to the EU, or to the third country must to be reported. In the case of blends of oils originating from more than one EU Member State or third country, one of the following mentions must be used: 'blend of olive oils of European Union origin' or a reference to the EU; 'blend of olive oils not of European Union origin' or a reference to origin outside the EU; 'blend of olive oils of European Union origin and not of European Union origin' or a reference to origin within the EU and outside the EU. An exception is the case where the olives were harvested in an EU Member State or third country other than that in which the mill where the oil was extracted is located. In this case, the designation of origin shall contain the following wording: *'(extra) virgin olive oil obtained in (the Union or the name of the Member State concerned) from olives harvested in (the Union or the name of the Member State or third country concerned)'*.

However, the regulation does not specify an official analytical procedure to verify the conformity of the label-declared geographical origin, and this has raised the interest of researchers to develop a reliable and effective method for purposes of authentication (Conte et al., 2019). During the last years, different analytical techniques have been applied in order to find potentially useful markers and efficient instrumental approaches that are able to discriminate olive oils according to their geographical origin.

In this regard, traditional chromatographic techniques, analyzing both major and minor compounds either individually or in a combined way, coupled or not with specific statistical chemometric data elaboration, have been investigated. A study in 2009 (García-González et al., 2009) proposed the application of artificial neural network (ANN) models for different levels of geographical classification (country, region, province, PDO) on a set of 687 EVOOs and VOOs from Spain, Italy, and Portugal, which were chemically characterized for the content of fatty acids, hydrocarbons, sterols, and alcohols. Other researchers evaluated the triacylglycerol (TAG) content and composition to discriminate Moroccan oils (Bajoub et al., 2016) and Croatian samples (Peršurić, Saftić, Mašek, & Kraljević Pavelić, 2018). In addition, the stereospecific distribution of fatty acids in TAGs was reported to be useful in discriminating olive oils from different areas of North-Eastern Italy (Vichi, Pizzale, & Conte, 2007). Specific metabolites such as sterols and phenolic compounds have been investigated to identify the optimal markers, and may be a promising approach to discriminate oils according to geographical origin (Giacalone, Giuliano, Gulotta, Monfreda, & Presti, 2015; Ben Mohamed et al., 2018; Ghisoni et al., 2019). Interesting findings have also been recently reported on sesquiterpene hydrocarbons as geographical markers (Quintanilla-Casas et al., 2020). Moreover, volatile compounds have been amply studied by applying different instrumental techniques combined with chemometric data elaborations (Kosma et al., 2017; Bajoub et al., 2018; Lukić, Carlin, Horvat, & Vrhovsek, 2019).

Furthermore, rapid and innovative instrumental approaches have been developed and tested in order to deal with the need for simple, rapid, and environmentally friendly techniques (Valli et al., 2016). This critical review (Valli et al., 2016) reports an overview of the principal applications of optical techniques (UV-Vis, NIR, MIR, RAMAN, NMR, and fluorescence spectroscopy), methods based on electrical characteristics, and instruments equipped with electronic chemical sensors (electronic nose and tongue) for discrimination of EVOOs and VOOs according to their geographical provenance. In addition to these approaches, other promising techniques include stable isotopes analysis (Angerosa et al., 1999; Chiocchini, Portarena, Ciolfi, Brugnoli, & Lauteri, 2016; Bontempo et al., 2019), multi-element fingerprint (Sayago, González-Domínguez, Beltrán, & Fernández-Recamales, 2018), differential scanning calorimetry (Mallamace et al., 2017), and GC-IMS (Gerhardt, Birkenmeier, Sanders, Rohn, & Weller, 2017).

Melucci and co-workers (Melucci et al., 2016) proposed the application of a Flash Gas Chromatography Electronic Nose (Heracles II) and a multivariate data analysis to control the compliance of information on geographic origin declared in the label ("100% Italian" vs "non-100% Italian") for the first time. This instrumental approach allows to realize the headspace analysis in short time and the results are processed by chemometric tools following an untargeted approach. For this reason, it can be considered as a fingerprint method, since the data can be elaborated for sample classification that is not aimed towards identification and quantification of specific analytes. Following these preliminary results and the actual need for a rapid and effective method for geographical authentication of VOOs, the aim of this work was the application of flash gas chromatography (Heracles II) for rapid discrimination of 210 EVOOs and VOOs according to geographical provenance. In this case, the categories considered for samples classification were EU member states vs third countries, and the data obtained were elaborated by applying two different classification techniques, one linear (Partial Least Square-Discriminant Analysis, PLS-DA) and one non-linear (Artificial Neural Network, ANN).

## 2. Materials and methods

### 2.1 Samples

A total of 210 EVOOs and VOOs with a different geographical origin were collected directly from companies that were also asked to provide, when available, information about location of the mill, type of plant used, olive variety, and commercial category (Table S1, Supplementary material). Considering that the indication of the geographical origin on the product label is mandatory for EVOOs and VOOs, samples belonging to both these two categories were included in this study.

According to geographical provenance, samples were distributed in 3 classes (Table 1): "EU" for oils coming from EU member states; "Extra-EU" for oils coming from third countries (outside EU); "Blends" for samples obtained by mixing oils coming from different EU state members or oils coming from EU state members and third countries.

Aliquots of each sample (50 mL) were stored at -18 °C in plastic dark bottles. Oil were defrosted for at least 12 h and stored at 12°C before analysis.

**2.2 Volatile compounds analysis by Flash Gas Chromatography**

The analysis of volatile compounds was carried out using the Flash Gas Chromatography Electronic Nose Heracles II (Alpha MOS, Toulouse, France). The instrument was equipped with two metal capillary columns (MXT-5: 5% diphenyl, 95% methylpolysiloxane, and MXT-1701: 14% cyanopropylphenyl, 86% methylpolysiloxane, for both columns: 10 m length, 180 µm internal diameter, 0.4 µm film thickness) working in parallel mode and different in polarity of the stationary phase. This permits slight differences in the separation capability of molecules detected by a FID applied at the end of each column.

Each sample was analysed in triplicate, weighing $2 \pm 0.1$ g of oil in a 20 mL vial sealed with a magnetic plug. For analysis, the vial was placed in a shaker oven for 20 min at 40 °C and 500 rpm. Next, 5 mL of the headspace were collected, introduced in a splitless injector (injector temperature 200 °C, injection speed 100 µL/sec, carrier gas flow, to ensure a fast transfer of the sample from the inlet to the trap, 30mL/min), and adsorbed on a Tenax® TA trap maintained at 40 °C for 60 sec to concentrate the analytes. The syringe temperature was set at 70 °C. Subsequently, desorption was obtained by increasing the trap temperature to 240 °C in 93 sec and the sample was injected (pressure of the carrier gas at columns' head 40 kPa.) and split (split flow 5 mL/min) into the two columns. The thermal program started at 40 °C (held for 2 sec), increased up to 80 °C at 1 °C/sec, and then to 250 °C at 3 °C/sec. Hydrogen was used as the carrier gas with a pressure from 40 kPa to 64 kPa, increasing with a rate of 0.2 kPa/sec. At the end of each column, a FID detector (detector temperature 260 °C) was placed and the acquired signal was digitalized every 0.01 sec. The software used to control the instrument was AlphaSoft version 14.5.

**2.3 Data processing**

For the data analysis, the full chromatograms were processed by applying chemometric elaborations with an untargeted approach. The raw data of each chromatogram (intensity values for each point of the chromatogram considering that the signal was digitalized every 0.01 sec, Palagano et al., 2019 [Dataset]) were exported from the software of the instrument and the data set with all the samples was imported into MatlabR2018a®. As data pre-treatment, chromatograms were aligned by COW (Correlation Optimized Warping) algorithm (Tomasi, Van Den Berg, & Andersson, 2004) and autoscaled (mean-centering followed by division of each column (variable) by the standard deviation of that column). Preliminary tests showed that chromatograms obtained from the MXT-5 column had a discriminant power higher than the other one (MXT-1701) and for this reason the classification models were developed considering only this column. Considering the reduced number of samples for the classes "Blend EU" and "Blend EU-Extra EU", these oils were grouped together with "EU" and "Extra-EU" samples, respectively. This means that for the data elaboration only two sample categories were considered: "EU" and "Extra-EU".

Two different statistical techniques were used to classify samples according to their geographical origin, the first (PLS-DA) based on a linear approach, and the second (ANN) on a non-linear approach.

In particular, the PLS-DA model was built using the PLS Toolbox for Matlab2018a®: intensity values of each point of the chromatogram, for a total of 19,900 data points, were used as variables X (matrix X), while the origin ("EU" and "Extra-EU") was implemented as variable Y (binary variables, 0 - 1). The sample data set was split into a calibration/full-cross validation set (75% of the sample) and an external validation set (25% of the sample) using the Kennard-Stone method (selects samples that best span the same range as the original data, but with an even distribution of samples across the same range) (Daszykowski, Walczak, & Massart, 2002). The threshold value useful to define the category of each sample was defined using a probabilistic approach based on Bayes's rule.

The ANN model was performed by using the Neural Net Pattern Recognition tool for Matlab2018a®. Specifically, a Multi-Layer Perceptron (MLP) neural network was built to predict the specific class to which samples belong using a non-linear method. For input and hidden layers, linear and logistic activation functions, respectively, were used, while for output layer the SoftMax function was applied. From a statistical point of view, with the SoftMax activation function and cross entropy error, the output is interpretable as posterior probabilities for categorical target variables (Bishop, 1995). One nominal output variable is returned, assuming that the target output is 1.0 in the correct class output, and 0.0 in the non-correct class. Looking for the best classification ability, different node numbers in the hidden layer and combinations were tested. The convergence of ANN was ruled by a back propagation algorithm. The original data set was randomly divided into a training set (60%), verification set (20%), and test set (20%). The training set was used to calculate the transfer function parameters of the network, the verification set to indicate possible over-learning, and the test set was treated as an unknown, the correct classification of which indicates that the neural network is performing well. It was checked that samples from both classes were contained in the test set.

## 3. Results and discussion

A set of 210 EVOOs and VOOs were analyzed for their volatile profile by flash gas chromatography. Considering the large amount of data and aim of this work, chemometric elaborations following an untargeted approach were carried out.

For elaborations, samples were grouped into two categories: "EU", that included oils from single EU state members and blends of oils from different EU countries, "Extra-EU" that consisted of oils from single countries outside the European Union and blends of oils from the EU and third countries.

In Figure 1-a the mean chromatogram of "EU" and "Extra-EU" categories, obtained averaging the intensity of each variable for all "EU" or "Extra-EU" samples, is reported: even if almost all peaks are concentrated in the initial part of the chromatogram (between 2000 and 10000 variables), a clear difference, in terms of variable intensities, exists between the two groups, thus confirming the discriminating power of the volatile profile with respect to the geographical origin (Melucci et al., 2016; Lukić, Carlin, Horvat, & Vrhovsek, 2019).

Concerning the PLS-DA results, the values of the estimated Y variable (geographical category) by the model in cross and external validations are shown in Figure 2. The dotted line identifies the threshold value used to define the attribution of samples to different classes. Regarding the location of each sample, a greater distance from the threshold line can be interpreted as a better classification capacity of the model.

The results, in terms of percentage and number of samples correctly classified, are reported in Table 2. The percentage ranged from 80.8% to 91.2%. The values obtained for the "EU" category were higher, likely because of the greater number and variability of samples used to build the model. The external validation percentages were lower compared to those obtained for the cross-validation as expected, but the results can be considered more robust since they were obtained considering the 25% of samples that were not used to build the model.

The VIP (Variable Importance in Projection) score obtained by the PLS-DA confirmed that the section of the chromatogram ranging from 2000 to 10000 variables has a major contribution to sample discrimination (VIP values greater than 1) according to geographical origin (Figure 1-b).

Focusing on those incorrectly classified samples, a specific trend as a function of characteristics that could usually affect the volatile profile of the oil (such as the commercial category, olive cultivar, or country of origin) was not seen.

Results related to the probabilistic approach are shown in Figure 3. The graph refers to the category "EU": this means that higher a sample is located, the higher the probability for which it is classified as member of the "EU" category. As a consequence, oils classified as members of the other category (Extra-EU) are located in the bottom area of the graph. In this case, the threshold value is fixed at 0.5, corresponding to a probability of 50%: a sample classified with a probability lower than this is considered as not correctly grouped. It is also interesting to note that most of samples were correctly classified with a probability between 90% and 100%.

Regarding ANN, an early stopping technique was used to select the number of training cycles (epochs) to avoid over-fitting, using the test set to monitor the prediction error. An example of this procedure is reported in Figure 4, where the best ANN training was characterized by 18 epochs. Above this point, the error increased further indicating that the ANN tends to overfit. Consequently, the results of ANN are related to these iterations.

Training was repeated 5 times and the network's predictions were averaged, since with ANNs convergence is influenced by the initial weight value and the randomized split of data in training, validation, and test sets. The best prediction results were obtained with a three layers network, having 5 nodes; a larger number of nodes did not increase the network performance.

The classification results, in terms of percentage of samples correctly classified, are summarized in Table 3. Means and standard deviations (in brackets) were taken into account.

As reported for the PLS-DA model, even in this case higher percentages (from 93.2% to 98.7%) were achieved for the "EU" category in all the three data sets.

Comparing the results of the external validation (PLS-DA) and testing (ANN), it is possible to note that higher percentages were obtained in the second case for both the "EU" and "Extra-EU" categories. In particular, an increment of 4.7% and 8% of samples correctly classified was obtained. This is probably due to the fact that the ANN model is based on a non-linear approach.

In general, the percentages obtained were slightly lower than those reported by other studies based on volatile compounds and chemometric untargeted data elaboration (Gerhardt, Birkenmeier, Sanders, Rohn, & Weller, 2017; Bajoub et al., 2018; Lukić, Carlin, Horvat, & Vrhovsek, 2019). This aspect can be explained by the great variability, in terms of geographical origin, olive variety, commercial category, of the samples analyzed, which represents a strong point of this work.

The results described herein confirm the suitability of flash gas chromatography for checking geographical traceability of EVOOs and VOOs, even using untargeted chromatographic signals of the volatile fraction as variables for multivariate analysis (Melucci et al., 2016). An in-house validation of this analytical method, carried out to verify that a repeatable and reproducible signal, with sufficient sensitivity to collect the valuable information from the samples, has been carried out which underlined the good performance of the technique; this will be discussed in more detail in a subsequent publication.

## 4. Conclusions

In this work, the application of flash gas chromatography for volatile compounds analysis combined with untargeted chemometric data elaborations (PLS-DA and ANN) to discriminate EVOOs and VOOs with different geographical origin was presented.

For both elaborations, satisfactory results, in terms of percentages of samples correctly classified, were obtained: PLS-DA (external validation) allowed classification of around 89% and 81% of "EU" and "Extra-EU" samples, respectively; for ANN (testing set) the percentages were equal 93.2% and 88.8%, respectively.

It is important to highlight that these promising results were achieved by analyzing a set of samples that are representative of the large variety of parameters (olive cultivar, country of origin, commercial category) that can describe olive oil product and affect its chemical characteristics. The results obtained herein sustained the use of multivariate chemometrics with untargeted detection of volatile compounds as a powerful tool to discriminate EVOOs and VOOs of different origin. Other studies have already reported that the analysis of volatile compounds is suitable for tracing the geographical origin of VOOs. However, the methodology proposed herein presents some advantages in comparison with other techniques generally applied for this analysis, as it is very rapid (only 200 sec are needed for each chromatographic run) and easy to use since no sample treatment is required.

**Funding**

This work was supported by the Horizon 2020 European Research project OLEUM *"Advanced solutions for assuring the authenticity and quality of olive oil at a global scale",* which has received funding from the European Commission within the Horizon 2020 Programme (2014–2020), grant agreement no. 635690. The information expressed in this article reflects the authors' views; the European Commission is not liable for the information contained herein.

**Acknowledgments**

The authors would like to thank the colleagues involved in the OLEUM project, from EUROFINS, IPTPO, ITERG, UZZK, and ZRS for collection and shipment of the samples.

**Declaration of interest**

None.

| Origin class | N | Country of origin |
|---|---|---|
| EU | 116 | 29 Spain, 25 Italy, 22 Croatia, 16 Greece, 12 Portugal, 12 Slovenia |
| Extra-EU | 70 | 42 Morocco, 21 Turkey, 6 Tunisia, 1 Chile |
| Blends | 24 | 12 EU blends, 12 EU/Extra-EU blends |

**Table 1.** Number of samples for each origin class considered and geographical origin. EU: oils from EU state members; Extra-EU: oils from countries outside the European Union; Blends: oils obtained mixing EU oils or EU and Extra-EU oils.

| Category | Cross validation | External validation |
|---|---|---|
| EU | 91.2% (93/102) | 88.5% (23/26) |
| Extra-EU | 91.1% (51/56) | 80.8% (21/26) |

**Table 2.** Percentages and number (in parentheses) of correctly classified samples for each category using the PLS-DA model. EU: oils from a single state member of European Union and oils obtained by mixing EU oils; Extra-EU: oils from a single country outside the European Union and oils obtained by mixing EU and Extra-EU oils.

| Category | Training (%) | Validation (%) | Testing (%) |
|---|---|---|---|
| EU | 98.7 (1.1) | 95.4 (3.9) | 93.2 (3.2) |
| Extra-EU | 94.4 (7.0) | 88.7 (7.8) | 88.8 (5.4) |

**Table 3.** Percentages (mean of 5 training of the model and standard deviation in parentheses) of samples correctly classified for each category using the ANN model. EU: oils from a single state member of European Union and oils obtained by mixing EU oils; Extra-EU: oils from a single country outside the European Union and oils obtained by mixing EU and Extra-EU oils.